\documentclass[acmlarge]{acmart}
\usepackage[utf8]{inputenc}

\usepackage{booktabs} 
\newcommand{\tttt}[1]{\texttt{#1}}
\usepackage{subcaption}

\usepackage[ruled]{algorithm2e} 

\SetAlFnt{\small}
\SetAlCapFnt{\small}
\SetAlCapNameFnt{\small}
\SetAlCapHSkip{0pt}
\IncMargin{-\parindent}

\acmJournal{TALLIP}
\acmVolume{0}
\acmNumber{0}
\acmArticle{00}
\acmYear{2018}
\acmMonth{7}
\acmArticleSeq{00}

\setcopyright{usgovmixed}

\acmDOI{0000001.0000001}

\received{July 2018}
\received[accepted]{XXX XXXX}

\begin{document}
\title{Basic concepts and tools for the Toki Pona minimal and
constructed language}
\titlenote{Description of the language and main issues;
analysis of the vocabulary; text synthesis and syntax highlighting; Wordnet synsets}

\author{Renato Fabbri}
\authornote{This is the corresponding author}
\orcid{0000-0002-9699-629X}
\affiliation{%
  \institution{Institute of Mathematical and Computer Sciences, University of São Paulo}
  \streetaddress{Avenida Trabalhador São-carlense, 400 - Centro}
  \city{São Carlos}
  \state{SP}
  \postcode{668}
  \country{BR}}
\email{renato.fabbri@gmail.com}

\begin{abstract}
A minimal constructed language (conlang)
is useful for experiments and comfortable for making tools.
The Toki Pona (TP) conlang is minimal both in the vocabulary
(with only 14 letters and 124 lemmas)
and in the (about) 10 syntax rules.
The language is useful for being a used and somewhat established
minimal conlang with at least hundreds of fluent speakers.
This article exposes current concepts and resources
for TP,
  and makes available Python (and Vim) scripted routines for
the analysis of the language,
synthesis of texts, syntax highlighting schemes,
  and the achievement of a preliminary TP Wordnet.
Focus is on the analysis of the basic vocabulary,
as corpus analyses were found.
The synthesis is based on sentence templates,
relates to context by keeping track of used words,
and renders larger texts by using a fixed number of phonemes (e.g. for poems)
and number of sentences, words and letters (e.g. for paragraphs).
Syntax highlighting 
reflects morphosyntactic classes given in the official dictionary
and different solutions are described and implemented
in the well-established Vim text editor.
The tentative TP Wordnet is made available in three patterns
  of relations between synsets and word lemmas.
In summary, this text holds potentially novel conceptualizations about,
and tools and results in analyzing, synthesizing and syntax highlighting the
TP language.
\end{abstract}

%
%


%
%



\maketitle

\renewcommand{\shortauthors}{Fabbri, R.}

\section{Introduction}\label{intro}
Toki Pona (TP) is a minimal conlang (constructed language)
with only 124 words (120 without the synonyms).
Therefore, the concepts are usually very general and different,
and, without context, the words are not frequently related through
meronymy and hyponymy.
Such a linguistic setting is desired because of the simplicity
which entails facilitated e.g. learning and tool making.
Another reason why the minimal language design is compelling
is the study and harnessing of the strong and weak forms of
the Sapir-Whorf hypothesis (linguistic relativity),
i.e. that language dictates or at least influences thought
and world experience~\cite{sapWho}.
Accordingly, one uses a conlang as a thinking tool (or platform)
or to make experiments about the influence language has on
the thoughts of the `speaker' (also writer and reader).
Toki Pona is often described as a tool to meditate,
to simplify the thinking processes, and as a way to modify
the mood and impressions about the world.~\cite{tpLang,kama,tp3,tp4}
In~\cite{interview}, Sonja Lang (the creator of TP),
describes that she has seen the language been used successfully
in the context of management, creation of texts, legal texts,
etc.

This article provides a conceptual overview of the language,
which is fit both to the newcomer and to the expert
for being considerably different
from what was found in the literature~\cite{tpLang,kama,memrise},
here with emphasis on simplicity and flexibility.
Novel software routines for analysis, synthesis,
syntax highlighting, and the achievement of a tentative TP
Wordnet~\cite{wordnet} are also herein presented.

Next subsections hold a description of the general resources available,
a historical note, and some words about natural and constructed languages.
Section~\ref{basics} describes the TP phonology and syntax.
Section~\ref{hacks} presents the software routines we made available
and immediate results, such as listings and statistics of words,
poems and short stories, coloring schemes, and preliminary TP Wordnet
versions.
Conclusions and further work are in Section~\ref{conc}.
Appendix~\ref{mytoki} holds considerations about the author's
usage of Toki Pona,
with thoughts about rule breaking and potentially new conlangs.
Appendix~\ref{ftp} holds final words in Toki Pona.

\subsection{Resources on Toki Pona}
One might categorize current resources for the Toki Pona
language in: references and learning material, corpus,
websites, interaction groups
(where users talk a post texts and comments), and software gadgets.
The main references of the language are:
  the official book ``Toki Pona: The Language of Good''~\cite{tpLang},
    authored by Sonja Lang, the creator of the language;
and the online book ``o kama sona e toki pona!'',
from jan Pije~\cite{kama}.
A number of tools and other resources for dealing with Toki Pona
were developed by the community.
The most complete list is supposedly~\cite{gdoc} and includes
videos, musical pieces, artistic texts, reference documents (e.g.
a Toki Pona - Esperanto dictionary), journalistic articles,
and software tools.
For a more comprehensive view of the resources available
for the user, we suggest following the links 
in~\cite{gdoc,wikiToki}.

\subsection{Historical note}
TP was developed as an internal and personal language
by Sonja Lang in late 1990s~\cite{interview}.
A description was released as a draft in 2001, and in 2007 some documents
reported it to have a few hundred speakers.
The English official book~\cite{tpLang} was published only in 2014.
In 2016, a version of the official book was published in French.
Nowadays, one finds a number of texts about TP and written
in TP (e.g. in social platforms such as Facebook groups,
microblogging, Telegram and IRC),
and other diverse uses of TP e.g. for artificial intelligence and software tools~\cite{gdoc}. 

\subsection{Natural and constructed and artificial languages}
A language one uses (or might use) to communicate by speaking
and writing is called `natural language'.
A `constructed language' (also planned or invented language),
e.g. Esperanto, Toki Pona, and Lojban, is a natural language built by someone or
a group.
An `artificial language' is most usually a term used
to designate a language yield by artificial agents,
such as by AI routines, or by humans in controlled experiments,
and are considered e.g. within `cultural evolution' studies.
Formal languages are defined by tokens and rules to
operate them, they span from computer programming languages
to math and formal models for natural languages.

Constructed languages (aka. conlangs) are in some contexts called artificial
languages, but creators most often prefer to use the term
`standardized', 'constructed'
or 'planned' language with the argument that the conlangs
are rooted on \emph{natural} languages,
and 'artificial' is thus misleading.
The preferred terms seem to be \emph{planned} or \emph{constructed} language
or simply \emph{conlang}.
The construction of languages is called glossopoeia.
TP is engineered for experiments, meditation and philosophy;
and useful at least as an auxiliary international language and
as an artistic language.~\cite{conlanWikip,wikiArtLang}


\section{Overview of the language}\label{basics}
This section describes very succinctly the formation of
words and sentences in the Toki Pona language.
It should enable a newcomer to grasp the essentials
of the language and the experienced to
acquire new insights.
Furthermore, it is a solid reference of the phonological and syntax
rules, and enables one to understand and modify the software presented
in Section~\ref{hacks}.

\subsection{Phonology}\label{phonology}
Words in Toki Pona are written using only 14 letters:
\begin{itemize}
  \item Vowels a (open), e (mid front), o (mid back), i (close front),
    u (close back).
  \item Consonants j, k, l, m, n, p, s, t, w:
    \begin{itemize}
      \item Nasal: m (labial), n (coronal).
      \item Plosive: p (labial), t (coronal), k (dorsal).
      \item Fricative: s (coronal).
      \item Approximant: w (labial), l (coronal), j (dorsal).
    \end{itemize}
\end{itemize}

There are standard guidelines for pronunciation,
but the language allows for considerable allophonic
variation.
For example, /p t k s l/ might be pronounced
[p t k s l] or [b d g z !R].
Especially for poetry, one might consider
j and w to be vowels
(e.g. j as 'i' and w as 'u').

Syllables are of the form (C)V(n):
an optional consonant, a vowel and an optional coronal nasal consonant
(n).
Non word-initial syllables must follow the pattern CV(N).
These sequences are forbidden: ji, wu, wo, ti, nm, nn.

\subsection{Syntax}\label{syntax}
\subsubsection{Fundamental notions}
As in other natural languages, 
colloquial TP might have
incomplete sentences and deviate from the
norm.
The basic structure of sentences is:
`subject' (Noun) li `predicate' (Verb) e `object' (Noun).
The li might be repeated to associate more than
one predicate to the subject.
The particle li is omitted if the subject is a simple mi (I or us)
or sina (you). A discussion about issues (potential problems)
yield by this last rule
and how the author deals with them is in Appendix~\ref{mytoki}.
The particle e might be repeated to associate more than
one object to a predicate.
Sentences might be related through la,
'sentence' la 'sentence', where the second sentence is
the main sentence, and the first sentence is a condition
to the second.
Multiple la-s are not described in literature,
but one might assume that the last sentences
are conditions to the next,
except in cases where the context suggests
otherwise.

Noun and verb phrases are (usually) built with the non-particle words.
The first word is the noun or verb and subsequent words
qualify the noun or verb (i.e. yield adjectives and adverbs).
The pi particle might be used to separate sequences of words
to be evaluated before the relation yield by pi.
As pi is often ill understood and used,
the following structures might be handy for newbies and as a
reference:
\begin{itemize}
  \item No pi, `word word word':
word $\leftarrow$ (qualifies 1) word $\leftarrow$ (qualifies 2)  word.
  \item One pi, 'word pi word word': word $\leftarrow$ (qualifies 2) [
      word $\leftarrow$ (qualifies 1)  word ].
  \item Two pi-s: `word pi word word word pi word word':
    word $\leftarrow$5 [word 2 word] 3 word $\leftarrow$4 word 1  word;
or:
    word $\leftarrow$5 [word 1 word] 2 word $\leftarrow$4 word 3  word.
\end{itemize}

\noindent Further notes on the usage of pi:
\begin{itemize}
  \item In a sequence of words, without pi, the second word qualifies
    the first, the third word qualifies the phrase yield by the first
    two words, the fourth word qualifies the phrase yield by the first
    three words and so on.
  \item It is redundant to use pi before the last word in a noun or
    verb phrase, reason why it is most often
    omitted.
    Its use in this case is considered an error~\cite{tpLang,kama},
    but, as one might notice, it does not add (much) information
    through syntax because the order of qualifications is conserved.
    It adds as an emphasis because of greater length of the segment, as a preparation: 'jan lili pi mama' (mommy's child).
    Also used in~\cite{akesiWawa}.
  \item The book by jan Pije~\cite{kama} describes another use for pi:
    after li to mean possession, e.g. `soweli li pi sina' (your pet).
    This employment of pi might thus be regarded as correct, but is promptly written
    as a noun phrase (e.g. `soweli sina') and is not mentioned
    by the official book~\cite{tpLang}.
\end{itemize}

All the words except the particles (li, e, la, pi, a, o, anu, en, seme, mu)
and the words classified solely as
prepositions (kepeken, lon, tan)
are usable after any noun or verb phrase,
(i.e. 107 words discarding synonyms).
Notice that the phrase expresses a noun (in a subject or
object) or a verb (in a predicate).
And that the first word of the phrase is the noun or verb,
and that subsequent words are (used as) adjectives or adverbs.
The pi precedes a noun or a verb to be qualified and then qualify
the phrase that comes before it.
These are reasonable expectations and not strict conventions.

At this point, the only missing syntax rule is related to
the prepositions: kepeken, lon, sama, tan, tawa.
They might appear at the end of a phrase,
might also be followed by another phrase,
and require no particle. E.g.
`toki *tan* jan Pije li pana e sona *tawa* mi'.
Because prepositions can be used as adjectives
or nouns or verbs or adverbs, etc, and might follow
any noun or verb phrase,
they are one of the four main sources of ambiguity:
small vocabulary;
prepositions; absence of li after sina and mi; incomplete sentences.
E.g. '(mi [li]) pana tawa kon e ilo pi suli mute'.

\subsubsection{Particles}\label{sec:par}
Beyond the structural particles (li, e, la, pi) presented
in last section, other particles are:

\begin{itemize}
  \item a or kin, emphasis.
  \item o, vocative or imperative ('jan lukin sitelen o, li wawa')
  \item taso, means however as sentence or 'only' if adjective.
  \item anu, en: 'or' and 'and'. Used for nouns in noun phrases when subject or after pi.
    For verbs, repeat li.
    For object nouns, repeat e.
    If the noun is complementing a preposition (e.g. tawa, lon),
    one might repeat the preposition.
    As TP is a recent language, and is particularly able to cope with
    variation due to its simplicity, one might advocate for
    using en and anu/en wherever there is no ambiguity.
    E.g.: 'mi tawa mute anu moku lon tenpo lili',
    in the official documentation,
    might otherwise be regarded as wrong in strict canonical TP.
  \item nanpa, denotes numbering.
  \item seme, for questions, used next to the thing being asked for.
    'tan seme la sina pana e sike?'.
  \item mu, for animal noises. It is not only a particle, as in the
    official dictionary, but also a noun and a verb:
    'mi pakala e luka. mi mu mute'.
\end{itemize}

The vocabulary specifies these morphosyntactic classes:
nouns, adjectives, verbs, pre-verbs, adverbs, prepositions, particles, and numbers.
Such classification helps the speaker, specially the newcomer, but
it also suggests a deviation from TP usage:
the words might be used indistinctly as a noun
(the first word of: a subject, a predicate when there is no object, an object,
or a prepositional complement),
an adjective (anything that follows an noun and is not a particle or a
preposition),
a verb (follows mi or sina or li),
or an adverbs (follows a verb).
The pre-verbs (wile, ken, awen, kama, lukin, sona),
might be followed by a verb, but might also be understood
as a verb qualified by the next word,
which carries a very similar if not identical meaning
('wile moku' might be understood both as a pre-verb followed
by a verb and as a verb followed by an adverb).
The pre-verb words are all
have other morphosyntactic classes,
in the official dictionary,
such as adjective, noun, and verb.
The only exception is wile, which is specified to be only a pre-verb,
but is trivially used as a verb:
'mi wile e moku e telo'.

Thus, the classes given in the dictionary dictate little
in practice:
jan kala li lape lon ni.
Where kala, lape and ni are in this phrase
as adjective, verb and noun,
and are in the dictionary as noun,
adjective and adjective.

\subsubsection{'li' as the verb 'to be'}
The particle li very often relates phrases as
the ver 'to be'. E.g. 'ona li pona' (that is beautiful)
and 'jan nasin li jan wawa' (someone with techniques is someone strong).

\subsubsection{Recognizing POS tags by speaker and machine}\label{sec:rec}
One might perform a reasonable POS tagging by following these rules:
\begin{itemize}
  \item Noun: the first word in a noun phrase.
    Starting a sentence if it is complete,
    after an 'e', after a pi.
    The predicate is often a noun
    if the sentence has no object.
  \item Adjective: second word on after an e and after
    second word on in the subject phrase if present.
  \item Verb: after a li, mi or sina.
    If there is no object, the predicate
    might be a noun or a verb.
  \item After a preposition, one might consider that what follows is 1) always
    a noun, or that 2) it can be a noun, adjective or verb,
    or that 3) it is a hybrid of noun, verb and adjective.~\cite{janKipo}
  \item If there is no object, the predicate might be parsed
as being a hybrid of noun, verb and adjective.
  \item Notice that there is ambiguity in the structure
    introduced by the omission of li after mi and sina.
    Also, when there is no object, a noun or a verb
    or an adjective might be in the verb position.
    These are sources of syntactic ambiguities
    in TP.
    They might be resolved or minimized considering the semantics
    of the words.
    An initial effort in this direction might be
    using the classes in the official dictionary to resolve
    ambiguities whenever possible.
    This solution is not optimal for correct POS tagging,
    and does not solve all possible ambiguities
    (there are words classified as noun and adjective,
    adjective and verb, noun and verb, particle and verb).
    Another source of ambiguity is the pre-verbs as described
    in the literature~\cite{tpLang,kama} and
    in the end of Section~\ref{sec:par}.
\end{itemize}


\subsubsection{Additional notes: synonyms, e ni, names, structure, other}\label{sec:add}
The only synonyms on Toki Pona are:
a and kin; lukin and oko; sin or namako;
ale or ali.
In formations such as
toki e ni:, wile e ni:, tan ni: etc.,
'(e) ni' might be omitted and : used alone.
Names are by default transliterated,
but might not be.
These and other issues are further described in Appendix~\ref{mytoki}.
John Clifford (a notable ``toki-ponist'', aka. jan Kipo)
states that there are not noun, verb, adjective (or even prepositional)
phrases in TP, but only phrases in a structure
yield by particles and content words~\cite{janKipo}.

Many other considerations might be made about the language,
such as additional (deprecated and new) words,
situations where li is used and avoided,
canonical and alternative employment of pi,
counting systems,
associations of TP word sequences to words in English,
usage of images to represent TP words and sentences,
and best practices in writing TP texts or for translation.
For these and other matters, visit~\cite{tpLang,kama,tp4,gdoc}.

\section{Software for analysis, synthesis, and syntax highlighting}\label{hacks}
In this section,
we describe software, statistic, natural language processing,
and visualization results:
1)
the analysis of the Toki Pona (TP) language through statistics
of the vocabulary obtained by processing the official dictionary;
2)
the synthesis of sentences, poems and paragraphs (e.g. short stories)
in TP;
3)
syntax highlighting for TP,
including fine-tuning and theoretical
considerations;
4)
an initial Wordnet~\cite{wordnet} of
TP words.
Corpus analysis was already found in~\cite{corpus}.
The Python scripts (i.e. the toolbox)
for obtaining all the results in this section,
and more, is publicly available in~\cite{tokipona},
to facilitate both the inspection of the results and the
generation of derivatives by other interested parties.
Such scripts are friendly to understand and alter.

\subsection{Statistics of the vocabulary}\label{sec:stat}
This section focuses on the statistics of the vocabulary
and of the syntactic rules:
the letters, phonemes, word sizes,
possible combinations for words and sentences.
The \tttt{TPTabFig} Python class 
was used to obtain all the measurements and tables
discussed herein,
and hold further measurements and sets of words
regarded as less suitable for this exposition than
for a deeper investigation.\cite{prv}

As described in Section~\ref{basics},
there are only 14 letters,
and phonemes are also very restricted.
There are 120 different words in the official vocabulary,
4 of them having synonyms.
A total of 124 tokens (or lemmas),
not counting proper nouns (names)
and punctuation.
Table~\ref{foobar} shows the number of words
related to each POS tag (or morphosyntactic class)
specified in the dictionary.

\begin{table*}[h!]
\begin{center}
\caption{"POS tags incident and chosen as preferential e.g. in text synthesis.
                    The official dictionary often relates tokens
                    to more than one POS tag.
                    For the text highlighting Plugin, for example,
                    a token has to have an established tag to have
                    a defined color.
                    On the Chosen column, the tokens were regarded only once
                    by choosing the first occurrence of ['PRE', 'VERB', 'PREPOSITION', 'PARTICLE', 'ADJECTIVE', 'NOUN', 'NUMBER'] in the official dictionary.}\label{foobar}
\begin{tabular}{  l | c   c  }
POS & All  & Chosen \\\hline
NOUN & 58  & 49 \\
ADJECTIVE & 40  & 34 \\
VERB & 15  & 13 \\
PARTICLE & 12  & 12 \\
PRE & 6  & 6 \\
PREPOSITION & 5  & 5 \\
NUMBER & 4  & 1 \\\hline
total & 140  & 120 \\
\end{tabular}\end{center}
\end{table*}

From the official 124 words, 
26 of them (20.97\%)
have only one syllable,
85 (68.55\%) have two syllables,
and 13 (10.48\%) have three syllables.
No official word has four or more syllables.
In all the 124 tokens, there are 235 syllables (68 different),
and Table~\ref{tab:freqSyl} exhibits the 10 most frequent
syllables in the first, last, middle and all positions.
Middle-word syllables only occur 13 times,
and are all different, with the exception of 'la'
which occurs twice.
A complete list of the syllables and their frequencies (in different
positions) is in the \tttt{tf.hsyls} variable after executing the \tttt{basic/tbBasic.py} script.
A list of all words, grouped by their size in syllables,
and ordered alphabetically and by the number of letters,
is in the \tttt{tf.hlsyl\_\_} variable.

\begin{table*}[h!]
\begin{center}
\caption{Frequency of syllables in Toki Pona
                    considering all 235 syllables of the 124 tokens,
                    only the first or last syllables or only the middle
                    syllables.
                    In parenthesis are the count and percentage of the
                    corresponding syllable. For more information and a
                    complete list of syllables, see Section~\ref{sec:stat}.
                    }\label{tab:freqSyl}
\begin{tabular}{  l | c   c   c   c  }
rank & all  & first  & last  & middle \\\hline
1 & li (13, 5.53\%)  & a (8, 6.45\%)  & li (10, 8.06\%)  & la (2, 15.38\%) \\
2 & la (10, 4.26\%)  & o (5, 4.03\%)  & lo (6, 4.84\%)  & je (1, 7.69\%) \\
3 & ka (9, 3.83\%)  & pi (5, 4.03\%)  & na (6, 4.84\%)  & ka (1, 7.69\%) \\
4 & na (9, 3.83\%)  & ka (4, 3.23\%)  & la (5, 4.03\%)  & ke (1, 7.69\%) \\
5 & pa (9, 3.83\%)  & la (4, 3.23\%)  & ma (5, 4.03\%)  & li (1, 7.69\%) \\
6 & a (8, 3.40\%)  & pa (4, 3.23\%)  & pa (5, 4.03\%)  & lu (1, 7.69\%) \\
7 & ma (8, 3.40\%)  & se (4, 3.23\%)  & ka (4, 3.23\%)  & ma (1, 7.69\%) \\
8 & si (8, 3.40\%)  & si (4, 3.23\%)  & sa (4, 3.23\%)  & me (1, 7.69\%) \\
9 & lo (7, 2.98\%)  & su (4, 3.23\%)  & si (4, 3.23\%)  & pe (1, 7.69\%) \\
10 & pi (6, 2.55\%)  & i (3, 2.42\%)  & te (4, 3.23\%)  & ta (1, 7.69\%) \\
\end{tabular}\end{center}
\end{table*}

Vowel and consonant frequencies,
and comparisons of vowels against consonants,
are as shown in Table~\ref{freqLet}
for start, end and internal positions.
The limited number of consonants favors
the language prosody to appear natural
as it resembles babbling.

\begin{table*}[h!]
\scriptsize
\begin{center}
\caption{Frequency of letters in Toki Pona. Columns
                    freq, freq\_I, freq\_L and freq\_M are
                    the frequencies of the letters in any, initial, last and middle
                    positions.
                    The columns 'v' and 'c' that follow them are frequencies
                    considering only vowels and consonants.
                    The most frequent vowel is 'a' in any position,
                    although it is more salient among words starting with a vowel
                    and among the last letter of the words.
                    For starting, ending and middle positions, the second most frequent
                    vowel varies.
                    Among the consonants, 'n' is the most frequent because it is
                    the only consonant allowed in the last position and because
                    almost 20\% of the words end with 'n'.
                    On the initial position, 's' is the most frequent consonant,
                    while in middle position 'l' is the most frequent consonant.
                    Many other conclusions may be drawn from this table and are
                    useful e.g. for exploring sonorities in poems.}\label{freqLet}
\begin{tabular}{  l | c   c   c | c   c   c | c   c   c | c   c   c  }
letter & freq  & v  & c  & freq\_I  & v  & c  & freq\_L  & v  & c  & freq\_M  & v  & c \\\hline
a & 16.35  & 33.19  & -  & 8.06  & 40.00  & -  & 29.03  & 35.64  & -  & 14.22  & 29.46  & - \\
e & 8.60  & 17.45  & -  & 2.42  & 12.00  & -  & 11.29  & 13.86  & -  & 10.78  & 22.32  & - \\
i & 11.53  & 23.40  & -  & 3.23  & 16.00  & -  & 20.97  & 25.74  & -  & 10.78  & 22.32  & - \\
o & 7.55  & 15.32  & -  & 4.03  & 20.00  & -  & 14.52  & 17.82  & -  & 6.03  & 12.50  & - \\
u & 5.24  & 10.64  & -  & 2.42  & 12.00  & -  & 5.65  & 6.93  & -  & 6.47  & 13.39  & - \\\hline
j & 2.10  & -  & 4.13  & 3.23  & -  & 4.04  & 0.00  & -  & 0.00  & 2.59  & -  & 5.00 \\
k & 6.29  & -  & 12.40  & 11.29  & -  & 14.14  & 0.00  & -  & 0.00  & 6.90  & -  & 13.33 \\
l & 9.22  & -  & 18.18  & 12.10  & -  & 15.15  & 0.00  & -  & 0.00  & 12.50  & -  & 24.17 \\\hline
m & 4.61  & -  & 9.09  & 10.48  & -  & 13.13  & 0.00  & -  & 0.00  & 3.88  & -  & 7.50 \\
n & 10.48  & -  & 20.66  & 6.45  & -  & 8.08  & 18.55  & -  & 100.00  & 8.19  & -  & 15.83 \\
p & 5.66  & -  & 11.16  & 11.29  & -  & 14.14  & 0.00  & -  & 0.00  & 5.60  & -  & 10.83 \\\hline
s & 6.29  & -  & 12.40  & 13.71  & -  & 17.17  & 0.00  & -  & 0.00  & 5.60  & -  & 10.83 \\
t & 3.14  & -  & 6.20  & 6.45  & -  & 8.08  & 0.00  & -  & 0.00  & 3.02  & -  & 5.83 \\
w & 2.94  & -  & 5.79  & 4.84  & -  & 6.06  & 0.00  & -  & 0.00  & 3.45  & -  & 6.67 \\
\end{tabular}\end{center}
\end{table*}

Within the rules exposed in Section~\ref{phonology}
(of 14 letters, 5 vowels, (C)V(n) phonemes,
forbidden ji, wu, wo, ti, nn, nm syllables),
96 words are possible with 1 syllable,
8256 with 2 syllables, 710016 with 3 syllables.
In the incident words of the official dictionary,
no middle syllable end with the nasal consonant 'n'.

Possible sentence structures are unlimited.
For a noun or verb phrase, one might quantify the possibilities
by considering all words but the particles that are not classified
also as something else
(i.e. li, e, la, pi, a, o, anu, en, seme, mu)
and the prepositions that are not classified also as something
else (i.e. kepeken, lon, tan).
This yields $120-13=107$ words.
As discussed in Section~\ref{syntax},
the author advocates that, although words have specific classes in the dictionary,
they might be used indistinctly as nouns, adjectives,
verbs and adverbs.
Thus, one has 107 possibilities of noun and verb phrases with
one word, $107^2=11,449$ possibilities with two words,
$107^3=1,225,043$ with three words and so on.
Be $n$, $v$, $o$, and $p$ the number of words in the subject (noun), predicate
(verb), object (noun) and prepositional (see Sections~\ref{sec:rec} and~\ref{sec:add}) phrases of a sentence,
and assume that the sentence has at most one prepositional phrase,
one might quantify the possibilities using the formula:
$\delta = 107^n\times 107^v\times 10^o \times 5\times 107^p$,
where $5$ stands for the possible prepositions.
To account for the particles, one possibility is
to assume for simplicity the use of at most one particle
(not li, e, la, pi, resulting in $8$ particles and $9$ possibilities)
at each phrase, yielding:
$\delta = 107^n\times 107^v\times 107^o \times 5\times 107^p\times 9^4$.
For example, assume $n=v=o=p=1$, then
$\delta=107^1\times 107^1\times 107^1 \times 5\times 107^1\times 9^4=
 4.300066\times 10^+12$,
i.e. more then 4 trillion possible
sentences with single word phrases (i.e. no adjectives or adverbs),
while allowing only one particle per phrase
and only one predicate phrase
(e.g. 'ona li moku e soweli lon supa').~\cite{tokipona}.

\subsection{Synthesis of text}\label{synth}
Such counting exercises are also useful
for (semi-)automated writing through scripting.
The syntax organizes the words in larger structures.
The rhymes are very restricted and sonorities are
bounded by the small vocabulary and simple syntax.
There are some specific tasks for achieving texts,
such as finding the number of syllables considering the elisions,
or handling interaction of the writer with the script
to choose sentences or verses or stanzas.
The package~\cite{tokipona}, herein presented, also
has capabilities for synthesizing TP text.
On its most basic level, these routines yield
noun, verb and prepositional phrases,
and sentences.
They also aim at making larger scale texts
by assuring the use of specific words (to entail context),
e.g. to assist the creation of short narratives,
and by employing stylistic outlines for poems.

\begin{figure}     \center
  \begin{subfigure}{.5\textwidth}
    \centering
        \includegraphics[width=.8\linewidth]{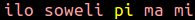}\vspace{-0.15cm}
          \caption{Synthesized phrase.}
            \label{fig:sfig1}
  \end{subfigure}\\\vspace{0.3cm}
  \begin{subfigure}{.9\textwidth}
    \centering
        \includegraphics[width=1\linewidth]{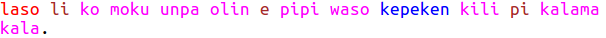}\vspace{-0.15cm}
          \caption{Synthesized sentence.}
            \label{fig:sfig2}
  \end{subfigure}\\\vspace{0.3cm}
  \begin{subfigure}{.8\textwidth}
    \centering
        \includegraphics[width=\linewidth]{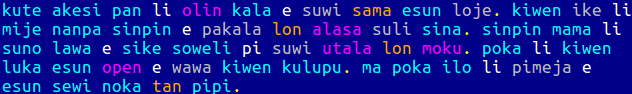}\vspace{-0.15cm}
          \caption{Synthesized paragraph.}
            \label{fig:sfig2}
  \end{subfigure}\\\vspace{0.3cm}
  \begin{subfigure}{.6\textwidth}
    \centering
        \includegraphics[width=.8\linewidth]{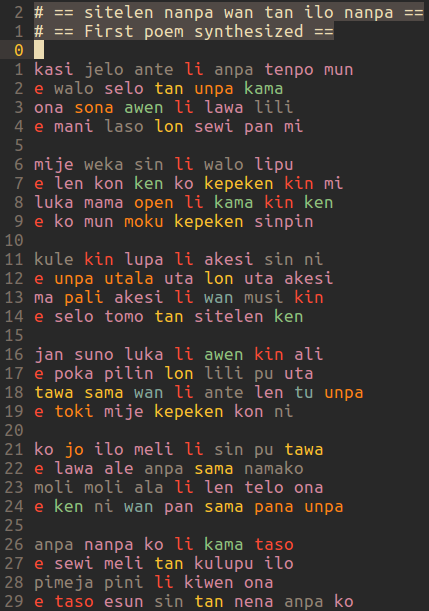}\vspace{-0.15cm}
          \caption{Synthesized poem.}
            \label{fig:sfig2}
  \end{subfigure}
  \caption{Toki Pona texts synthesized by an instantiated \tttt{TPSynth} Python class~\cite{tokipona}: (a) a phrase, (b) a sentence, (c) a paragraph,
  and (d) a poem.
  The unusual word combinations are convenient for exploring semantic
  possibilities.
  One might obtain many texts to select the excerpts
  she finds fit for the intended use.
  For more information, see Section~\ref{synth}.
  The words are colored in accordance with the considerations
  in Section~\ref{shigh}.}
  \label{fig:syn}
\end{figure}

Figure~\ref{fig:syn} holds some synthesized texts
with different color schemes for the syntax highlighting,
as presented in the next section.
The textual synthesis described here,
and implemented in the \tttt{TPSynth} class,
might be enhanced in countless ways,
e.g. as described in Section~\ref{conc}.

Automatic and randomized synthesis of texts in TP is particularly
useful because of the reduced vocabulary where each word is related
to a broad semantic field.
Ideas often make sense in unexpected ways, and thus
the synthesis yields a procedure to 
explore the semantic possibilities within TP.
One might object that the resulting texts are unusual 
and even consider that they often hold insubstantial or unsound meaning.
I advocate that these unexpected formations are desirable
for exploring the possibilities of the language and of thought,
and for artistic endeavors.
Also, to obtain texts which one finds usual or satisfactory
in more strict or personal terms, such a person might just write
them normally or use the synthesized texts as raw material.



\subsection{Syntax highlighting}\label{shigh}
The same package~\cite{tokipona}
has a Vim~\cite{vim} syntax highlighting plugin
for Toki Pona.
The Python scripts hold
routines to arrange the coloring schemes
(i.e. to specify which tokens
are related to which coloring groups)
which are stored in the \tttt{syntax/tokipona.vim} file.
An online syntax highlighting gadget is found at~\cite{tpNetSH},
and the solution here described presents a number of enhancements:
it is capable of coloring all morphosyntactic classes;
it might be fine-tuned in the colors and their relations to sets of
tokens; it is designed to be used within Vim
and might be exported as HTML; the highlighting scheme is promptly
rendered by a script; the resulting script might be used in combination
with virtually any color scheme.

Basically, the resulting syntax highlighting
distinguishes the words among the morphosyntactic
classes according to the official dictionary as given in
Table~\ref{foobar}.
Also, some classes might be further defined (e.g. words beginning with
vowels) or joined (e.g. distinguishing only particles against rest,
or particles and prepositions against the rest).
The colors are also promptly changed according to~\cite{vim}
and exemplified in the package documentation,
specially the \tttt{syntax/tokipona.vim} syntax file

Currently, the Python package synthesizes the
syntax file through the\\ \tttt{TPSynHigh} class.
The user has control of class precedence,
merging and further details through tweaking such routine.
The choice of precise coloring schemes
involves fine tuning the color scheme being
used in Vim (such as 'blue', 'elflord' and 'gruvbox'),
and Vim's highlighting schemes as described in~\cite{vim}.
The usage of the package and plugin might be performed,
in summary, through the following actions:
\begin{itemize}
  \item Installation of the plugin, so that
    syntax highlighting of TP texts will be performed in any \tttt{*.tp} or \tttt{*.tokipona} file.
  \item Tweaking of the syntax file by hand. Or
  \item Instantiating the \tttt{TPSynHigh} class and then executing the \tttt{mkSynHighFile()} method
    of the resulting object.
    This generates a new \tttt{tokipona.vim} syntax file, at the right location in Vim folders,
    according to object settings.
  \item Write a file with the \tttt{.tokipona} or \tttt{.tp} extension
    (inside Vim) using Toki Pona words.
    Reload the highlighting scheme using \tttt{:e} whenever you
    change the syntax file by hand or through the Python routines.
  \item Access the used highlighted groups with \tttt{:syntax}.
    Access all the highlighting groups with \tttt{:so \$VIMRUNTIME/syntax/hitest.vim} or \tttt{:hi}.
    Change the coloring of a set of terms by associating
    a used group (e.g. \tttt{tpADJECTIVE}) to an existing group (e.g.
    Visual) such as in \tttt{:highlight link tpADJECTIVE Visual}.
    This may also be achieved changing \tttt{TPSynHigh().colors}
    attribute before executing \tttt{mkSynHighFile()}.
\end{itemize}

\noindent Such procedures are facilitated by the \tttt{basic/tpBasic.py}.
One may change: associations of tokens to HIGs;
associations of regex patterns to HIGs;
associations of HIGs to other HIGs;
whole color schemes or just sections of it.
The last of these tasks is promptly performed using 
the \tttt{space-c-s} and \tttt{space-c-c} normal Vim commands
supplied by the color plugin of the PRV framework~\cite{vim}.
The other tasks are more easily performed by
scripting Vim, Python import and package source code,
depending on the user ability and intended tasks.
The use of different color schemes within the same syntax
file is exemplified in Figure~\ref{fig:syn}.
By copying only the \tttt{syntax/tokipona.vim} and \tttt{ftdetect/tokipona.vim} files,
a user might benefit from the standard Vim commands.
E.g. \tttt{:colorscheme blue},
\tttt{:colorscheme solarize}, \tttt{:colorscheme gruvbox} or
\tttt{:colorscheme elflord} to see the same text colored
with different color schemes (association of colors to sets of tokens).
To interfere directly on the colors chosen,
\tttt{:highlight Normal guifg=\#00000 guibg=\#0000ff}
will change the standard foreground (text) color to pure black
and background to pure blue.
See \tttt{:h gui-colors} and \tttt{:h highlight}
for the way you might edit colors directly.
One might obtain an HTML file with the colored TP text
using the \tttt{:TOhtml} command.


\subsubsection{Advanced syntax highlighting considerations}
Standard guidelines for syntax highlighting
depend heavily on cultural and use factors
and have scarce scientific studies~\cite{stack}.
There are informed projects such as Solarized~\cite{solarized},
which present solutions for some contexts.
Strikingly, standard guidelines for syntax highlighting were not found.
Therefore, we considered current data visualization
theory~\cite{dv1,dv2,dv3,dv4} to glimpse at the potential guidelines:
\begin{itemize}
  \item The use of blue and other high frequency colors (such as
    violet) for the background and the avoidance of their use
  to fill small objects, such as letters and words.
  \item Explore simplicity and elegance in the  coloring schemes.
    Most tokens should be preferably of the same color, i.e.
    a small number of colors might be explored to achieve a clean
    visualization of texts.
  \item A power-law distribution of tokens among colors might be
    well-suited to mimic natural occurring phenomena.
    The TP dictionary morphosyntactic classification of words
    resemble a long-tailed distribution: as shown in Table~\ref{foobar},
    few classes hold the vast majority of words.
  \item Physical stimuli might be related to perceptual stimuli both 
    through a power-law or an exponential law (respectively known
    as Steven's law and Weber-Fechner's law).
    This might be useful e.g. for coloring sets of tokens
    considering their similarity, or correctly choosing values
    for e.g. the hue or luminosity channels.
  \item One usually wishes to maximize contrast,
  although taste and less wearing combinations might
  also dictate the coloring choices.
  \item Stipulate axes of parameters to set colors.
  Wavelength or frequency is an obvious axis given the considerations
    previously exposed, but they are not perceptually uniform.
    Considering perceptually uniform color spaces in making
    choices may be relevant, such as CIElab and CIEluv.
  \item Consider two types of color schemes: of those with a dark background,
  which are more comfortable at first; and of those with a light background,
  which are usually impressive (and even annoying) at first,
  but the eye adapts in a few minutes, keeps you more stimulated,
    and is suitable for bright contexts (e.g. in daylight) .
\item The blue color is specially related to physiological
  stimulation of the body~\cite{blue,blue2},
    and programmers often report that a blue background keeps them
    awake and more concentrated (the author also notices such effect).
  \item Colors have been associated to enhancements in specific tasks,
    e.g. blue and red are respectively associated to enhanced
    creativity and detail-oriented tasks in~\cite{blue}.
  \item It is important to consider that \emph{a text editor user} might 
    stay many hours using the tool (and they very often do),
    and that the colors and formats involved are
  thus prone to entail a considerable effect in the body and mental
    activity, the quality of life, and work results of a writer
    (e.g. a programmer).
  \item Ideally, one should have facilities for tuning 
    the syntax highlighting (e.g. through keyboard shortcuts)
  as envisioned and implemented in~\cite{vim}.
\end{itemize}

The considerable irrelevance of the morphosyntactic classes,
described in Section~\ref{syntax}, suggests that an appropriate coloring
of words should
either distinguish only between particles, prepositions and the rest,
or consider the syntax to identify the nouns, verbs, adjectives, adverbs,
prepositions and particles.
Also, the coloring of poems might be more appealing if considered
e.g. the counting of syllables, the repeated letters or syllables,
and the ending syllables.
Further insights might be obtained through the
choices made and advocated in software such as text editors
(e.g. Vim, Emacs, Sublime),
packages dedicated to syntax highlighting (e.g.
pigments\footnote{\url{http://pygments.org/docs/}} and 
highlight.js\footnote{\url{https://highlightjs.org/}}),
and
other software (e.g. Linux terminals such as Xterm and gnome-terminal).
Finally, the careful choice of fonts is known to have
a relevant impact in comfort and productivity~\cite{fonts},
e.g. sans-sheriff fonts are more promptly read and yield a cleaner
text than a serif font, and the same applies to a monospaced font
which is more likely to yield a cleaner text, at least for programming.
One might also 
consider the mapping the textual structures
to sound~\cite{mass} (i.e. parametrize the synthesis of sounds e.g. 
by the counting of specific tokens inside a script, a function or
class or around a variable of conditional or
loop), which might be called ``syntax sonification''.

\subsection{Toki Pona Wordnet}\label{wn}
A first Toki Pona Wordnet was constructed
relating each of the TP words in the dictionary
to English Wordnet synsets~\cite{wordnet}
through the English lemmas.
The canonical (i.e. Princeton) Wordnet only contains nouns,
adjectives, verbs, and adverbs.
Thus, particles were not considered.
Numbers were considered adjectives.
Words presented as adjectives in the dictionary
were considered both as adjectives and adverbs.
Prepositions were considered in all classes.~\cite{wordnet}

The \tttt{TPWordnet} class~\cite{tokipona},
provides such tentative TP Wordnets in their simplest form:
the TP words are keys in a dictionary that returns the
corresponding synsets.
Three of such dictionaries are implemented:
one where all synsets related to the lemma is
returned (the version most consistent
with the expositions in Section~\ref{basics}),
with 4,027 synsets;
another as such but excluding the prepositions,
with 3,929 synsets;
a last one that returns only the synsets
registered in Wordnet with the same POS tag as the lemma
is in the official dictionary,
with 2,462 synsets. 
Another immediate possibility, not implemented,
would be to relate each Toki Pona word to the synsets
simultaneously associated to all the English
words bounded by a semicolon in the dictionary.
Such collection of synsets, and its relation to the whole
Wordnet
(the total number of synsets in Wordnet is $117,659$),
might guide creation of other conlangs
(e.g. one might seek to use lemmas related to synsets that
are very far apart, or that has the most complete neighborhood
possible).

TP words are very general
and each of them might mean many things.
Thus, the words are not very easily related e.g. by hyponymy
or meronymy.
Exceptions: jan (people) is a hyponym of soweli (mammal);
kili is an hyponym of kasi;
walo, pimeja, jelo, loje, laso
are hyponyms of kule.
There is at least one caveat:
if a particular meaning of each word is chosen,
then there might be many other of such relations,
e.g. 
  lawa, luka, noka, sinpin, monsi, lukin, mute, kute, palisa, and lupa
  might be considered meronyms of sijelo.
  Antonyms are also specially useful in TP: suno and mun, pona and jaki or ike, sinpin and munsi, lawa and noka, mije and meli, sike and palisa, pana and kama jo, pimeja and walo, weka and poka, sama and ante, ali and ala, anu and e, selo and insa. 

\section{Conclusions and further work}\label{conc}
This document presented a potentially novel description of Toki Pona (TP)
language in Section~\ref{basics}, and innovative and useful software routines for dealing with
the language in Section~\ref{hacks}.
As a minimal conlang, TP is very convenient for cognitive
experiments
through linguistic semantics and
for devising tools for analysis, creation and visualization,
a fact to which this document is evidence.
Other conlangs, and even non-planned languages might benefit from
TP and this content in many ways, e.g. one might
synthesize TP texts and translate them as convenient,
adapt the routines to obtain statistics of the vocabulary,
or take advantage of the language description to devise new conlangs
or stylistic guidelines for any language.

At the moment, the framework in~\cite{prv} is being used, by the author, for TP chatter bots
and for semantic explorations of domain knowledge.
Some of the many potential next steps should be highlighted:
\begin{itemize}
  \item The text synthesis facilities described in
    Section~\ref{synth} might be enhanced by further employing e.g.
    rhythm, meter, rhyme and form techniques~\cite{wikiPoetry}.
  \item Syntax highlighting that colors the tokens with respect
    to the syntactic positions and functions, and not in a fixed manner
    as is implemented and exposed in Section~\ref{shigh},
    may be implemented by using n-grams and
    other techniques from Natural Language Processing.
  \item The TP Wordnet(s) described in Section~\ref{wn} should be may
    be enhanced:
    implement a NLTK-like Wordnet interface; double-check if each synset
    is correctly associated to the TP words; seek further synsets
    that were not found by the English lemmas in the official
    dictionary; implement the most restricted version described in
    Section~\ref{wn}, which relates each TP word only to the synsets
    that are related to all the English words not separated by a
    semicolon; scrutinize the neighborhood of the synsets of the TP
    Wordnet in the English Wordnet.
  \item Conlanging: use Wordnet for making new conlangs;
    use insights derived from TP Wordnet.
    Create conlangs for specific uses:
    describing data, programs,
    scientific writings, creative writing
    (exploring the thinking process).
    A conlang might have different modes that are
    specified by section headers or tags. 
  \item Explore TP relation to other languages
    (e.g. suno and suwi might come from sun and sweet),
    and the reasons that led Sonja (and maybe other people)
    to choose the 14 letters and the syllable structure.
    This might require a dedicated communication with the
    speaker community and the documentation keepers.
  \item Understand how the corpus was gathered in \url{tokipona.net}.
  \item Corpus-based analysis.
  \item Publication of original TP texts and translations.
  \item Write academic articles in TP
    (extending Section~\ref{ftp} which might be the
    first section of a scientific document written in TP).
    The outline might be:
    a summary in both English and Toki Pona, for
    facilitated acquisition of context,
    and an article in a canonical structure
    (e.g. introduction and related work, materials and methods,
    results and discussion, conclusions and further work),
    with an Appendix explaining the content and context in English.
    Maybe something about complexity, statistics, physics,
    or computer science;
    or linguistics, philosophy, literature
    or psychology.
    Partners may learn the language in a few hours or weeks
    (with or without help).
    This is a very challenging step.
  \item Make TP variations where each TP word is replaced by a word in
    a language with a large speaker population. 
For example,
suno might be sun in an English variant,
and sol in a Portuguese variant.
A naive version might be obtained through
choosing the first word in the description of
an official vocabulary, as they already exist in English and French.
  \item Explore a minimal conlang to describe software
    routines.
    In this sense, it seems advantageous
    to understand what `non-ambiguous' means in Lojban,
    and in what sense one is able to compile and parse Lojban.
    Consider the possibilities of and reasons for parsing TP.
\end{itemize}

\appendix
\section{Current usage of Toki Pona by the author}\label{mytoki}
Disclaimer: this whole section holds personal views and usages of
the Toki Pona (TP) language.
It is reasonable to use the standard sounds, but often [z] instead of s
is more comfortable.
Translation into Toki Pona (e.g. of biblical excepts),
and the creation of new texts as poems and short stories,
is both challenging and pleasant.
Texts are shared e.g. in~\cite{tokisona}
and/or published in the TP Facebook groups.
The li particle is most often not used after subjects sina and mi,
in accordance with the norm,
but, even so, they are useful when there are many predicates.
E.g. sina li wawa li pimeja li lukin pona li moku e kasi mute.
In such cases, the first li is sometimes omitted.
Also, li is found after mi and sina where one finds
that there is unwanted ambiguity, e.g.
sina moku pona 
which 
might be 'sina li moku pona' or 'sina moku pona (li jo e sike)'.

Names are by default transliterated,
but one should notice that, as in other languages,
names might be used as they are in the
correspondent mother tongue.
E.g. the name Erdös is used in
English and Portuguese although the standard
alphabet does not contain ö in such languages.
The use of e.g. English (or German) words
in TP texts is found when needed,
as happens often in scientific writing
(kernel is a German word used in English,
webpage and software are English words used in Portuguese).

Proposed notations for numbers seem numerous.
E.g. one might indicate if two numbers
are being multiplied (pi) or are in different scales
(such as in decimal or binary notations):
luka (pi) two mean 52.
Most importantly,
52 is a reasonable notation for a simple language.
E.g.
'mi jo e jan sama nanpa 12',
or
'ona li lon e soweli 27'.

Avoiding needless words...
'e ni:' may most often be written as ':'.
Subject phrase may be omitted (sometimes also the li)
if it is the
same as in the last sentence.
One may vary TP by grouping e.g. these words:
noka and anpa; luka, suno, sike and lawa; pali and pana.

Also, ambiguities introduced by omission of li
and absence of a token to denote preposition,
suggest the possibility of a syntax that is always uniquely
parsed (or at least less ambiguous).

It is tempting to conceive a language with the same syntax of TP:
'sentence' la 'sentence = subject + predicate + object' (each term with
a possible prepositional complement),
but always using particles to bound the sentence sections,
repeating them when the function is performed in the smallest scale:
'toki pi toki pona pi jan sona' meaning
language in toki pona and of intellectuals vs
'toki pi toki pona pi pi jan sona'
meaning language in [toki pona of intelectuals].
The keyword for initiating a preposition complement 
might be 'a', but it is already taken in TP.
One possibility is to use 'a' before a preposition and use 'ha' or 'he'
instead of TP a.
In such a setting,
one might enable pi, li, e, a (concept-qualifier, subject-predicate,
predicate-object, concept-preposition)
in any slot of the syntax template.
Between sentences one may use:
la pi, la li, la e, la a, and la la.
This is a very fractal conlang proposal.
Maybe also have a way to discern between a noun,
a qualifier (adjectives/adverbs) and a verb,
and accept any of them for the subject, predicate, object, preposition,
slots. Or assume a part-of-speech as default for a slot or for the
first word in a slot.

\section{Final words in Toki Pona}\label{ftp}
toki li nasin e lawa.\\
li nasin tawa (pi) toki insa.

toki pona li pona e nasin tan ni:\\
ona li jo e nimi nanpa lili.\\
ona li pona.\\
pona kepeken weka 'p' li ona, a.

o taso la toki pona li kalama li lukin pona.\\
sitelen en nasin li open e sitelen\\
suli~\cite{tpLang,kama,akesiWawa,gdoc,tokisona,Wikipesija}.\\
li sona. li nasin e toki e sona e lawa e lon.

toki ni li wawa tawa jan mute nasin en toki.\\
wawa tawa toki pi jan sona.\\
pi ilo nanpa en nanpa nasin.\\
taso tawa toki sona a.\\
sitelen sona, sitelen musi.

ilo lon Poki~\ref{hacks} li pana e sitelen\\
e sona tan sitelen,\\
en nasin.\\
Poki~\ref{basics} li pana e nasin pi toki pona.\\
e sitelen tawa kama sona.

mi wile pali e sitelen lon toki pona.\\
sitelen lon nasin, sitelen sona,\\
sitelen musi.\\
ante e toki pona\\
la ona li nasa.\\
taso nasa li pona mute.\\
li pona mute tawa lawa,\\
tawa kama sona, tawa sitelen e toki.\\
ante la toki pona o.\\
toki e toki pona tawa sina.\\
kama sona e toki pona sina.

o pona tawa jan pi toki pona.\\
tawa jan Sonja, Birns-Sprage, Kipo, Pije,\\
Siwejo, Malija, Tepan, jan kulupu mute.

\begin{acks}

The author thanks
FAPESP (project 2017/05838-3, coordinated by Prof. Dr. Maria Cristina Ferreira de Oliveira);
Toki Pona, Python and Vim authors, developers and literature maintainers;
Toki Pona user and speaker communities; 
Mario Alzate Lopez for introducing me to Toki Pona,
and Silverio Guazzelli Donatti for the stimulus that resulted
from the usage of Toki Pona for chatting.

\end{acks}

\bibliographystyle{ACM-Reference-Format}
\bibliography{sample-bibliography}


\begin{thebibliography}{32}


\ifx \showCODEN    \undefined \def \showCODEN     #1{\unskip}     \fi
\ifx \showDOI      \undefined \def \showDOI       #1{#1}\fi
\ifx \showISBNx    \undefined \def \showISBNx     #1{\unskip}     \fi
\ifx \showISBNxiii \undefined \def \showISBNxiii  #1{\unskip}     \fi
\ifx \showISSN     \undefined \def \showISSN      #1{\unskip}     \fi
\ifx \showLCCN     \undefined \def \showLCCN      #1{\unskip}     \fi
\ifx \shownote     \undefined \def \shownote      #1{#1}          \fi
\ifx \showarticletitle \undefined \def \showarticletitle #1{#1}   \fi
\ifx \showURL      \undefined \def \showURL       {\relax}        \fi
\providecommand\bibfield[2]{#2}
\providecommand\bibinfo[2]{#2}
\providecommand\natexlab[1]{#1}
\providecommand\showeprint[2][]{arXiv:#2}

\bibitem[\protect\citeauthoryear{authors}{authors}{2016}]%
        {stack}
\bibfield{author}{\bibinfo{person}{Various authors}.}
  \bibinfo{year}{2016}\natexlab{}.
\newblock \bibinfo{title}{Syntax-highlighting color scheme studies.}
\newblock \bibinfo{howpublished}{Software Engineering at Stack Exchange}.
\newblock
\urldef\tempurl%
\url{https://softwareengineering.stackexchange.com/questions/89936/syntax-highlighting-color-scheme-studies}
\showURL{%
Retrieved July 01, 2018 from \tempurl}


\bibitem[\protect\citeauthoryear{Blahu{\v{s}}}{Blahu{\v{s}}}{2011}]%
        {tp3}
\bibfield{author}{\bibinfo{person}{Marek Blahu{\v{s}}}.}
  \bibinfo{year}{2011}\natexlab{}.
\newblock \showarticletitle{Toki pona--eine minimalistische Plansprache}.
\newblock \bibinfo{journal}{\emph{Spracherfindung und ihre Ziele. Beitr{\"a}ge
  der}}  \bibinfo{volume}{20} (\bibinfo{year}{2011}), \bibinfo{pages}{51--56}.
\newblock


\bibitem[\protect\citeauthoryear{community}{community}{2018a}]%
        {gdoc}
\bibfield{author}{\bibinfo{person}{Toki~Pona community}.}
  \bibinfo{year}{2018}\natexlab{a}.
\newblock \bibinfo{title}{toki pona central hub preparation}.
\newblock
\newblock
\urldef\tempurl%
\url{https://docs.google.com/document/d/1Dzs-imNeZ8TMgdHUiiungJ4Yf97CJk9ylhQPXjWLsJU}
\showURL{%
Retrieved July 01, 2018 from \tempurl}


\bibitem[\protect\citeauthoryear{community}{community}{2018b}]%
        {memrise}
\bibfield{author}{\bibinfo{person}{Toki~Pona community}.}
  \bibinfo{year}{2018}\natexlab{b}.
\newblock \bibinfo{title}{Various Memrise courses on Toki Pona}.
\newblock
\newblock
\urldef\tempurl%
\url{https://www.memrise.com/courses/english/?q=toki+pona}
\showURL{%
Retrieved July 01, 2018 from \tempurl}


\bibitem[\protect\citeauthoryear{community}{community}{2018c}]%
        {Wikipesija}
\bibfield{author}{\bibinfo{person}{Toki~Pona community}.}
  \bibinfo{year}{2018}\natexlab{c}.
\newblock \bibinfo{title}{Wikipesija}.
\newblock
\newblock
\urldef\tempurl%
\url{http://tokipona.wikia.com}
\showURL{%
Retrieved July 01, 2018 from \tempurl}


\bibitem[\protect\citeauthoryear{et~al.}{et~al.}{2017}]%
        {janKipo}
\bibfield{author}{\bibinfo{person}{John~Clifford et al.}}
  \bibinfo{year}{2017}\natexlab{}.
\newblock \bibinfo{title}{o awen, akesi wawa}.
\newblock
\newblock
\urldef\tempurl%
\url{https://www.facebook.com/groups/sitelen/permalink/1597077093680003/}
\showURL{%
Retrieved July 01, 2018 from \tempurl}


\bibitem[\protect\citeauthoryear{Fabbri}{Fabbri}{2017a}]%
        {vim}
\bibfield{author}{\bibinfo{person}{Renato Fabbri}.}
  \bibinfo{year}{2017}\natexlab{a}.
\newblock \showarticletitle{An anthropological account of the Vim text editor:
  features and tweaks after 10 years of usage}.
\newblock \bibinfo{journal}{\emph{arXiv preprint arXiv:1712.06933}}
  (\bibinfo{year}{2017}).
\newblock


\bibitem[\protect\citeauthoryear{Fabbri}{Fabbri}{2017b}]%
        {akesiWawa}
\bibfield{author}{\bibinfo{person}{Renato Fabbri}.}
  \bibinfo{year}{2017}\natexlab{b}.
\newblock \bibinfo{title}{o awen, akesi wawa}.
\newblock
\newblock
\urldef\tempurl%
\url{https://www.facebook.com/groups/tokiponataso/permalink/1895942817391318/}
\showURL{%
Retrieved July 01, 2018 from \tempurl}


\bibitem[\protect\citeauthoryear{Fabbri}{Fabbri}{2018a}]%
        {prv}
\bibfield{author}{\bibinfo{person}{Renato Fabbri}.}
  \bibinfo{year}{2018}\natexlab{a}.
\newblock \bibinfo{title}{PRV: Python, RDF and Vim}.
\newblock
\newblock
\urldef\tempurl%
\url{https://github.com/ttm/prv/}
\showURL{%
Retrieved July 03, 2018 from \tempurl}


\bibitem[\protect\citeauthoryear{Fabbri}{Fabbri}{2018b}]%
        {tokipona}
\bibfield{author}{\bibinfo{person}{Renato Fabbri}.}
  \bibinfo{year}{2018}\natexlab{b}.
\newblock \bibinfo{title}{Toki Pona tools and article}.
\newblock
\newblock
\urldef\tempurl%
\url{https://github.com/ttm/tokipona}
\showURL{%
Retrieved July 01, 2018 from \tempurl}


\bibitem[\protect\citeauthoryear{Fabbri}{Fabbri}{2018c}]%
        {tokisona}
\bibfield{author}{\bibinfo{person}{Renato Fabbri}.}
  \bibinfo{year}{2018}\natexlab{c}.
\newblock \bibinfo{title}{Toki Sona: a blog dedicated to Toki Pona}.
\newblock
\newblock
\urldef\tempurl%
\url{http://tokisona.github.io/}
\showURL{%
Retrieved July 01, 2018 from \tempurl}


\bibitem[\protect\citeauthoryear{Fabbri, Junior, Pessotti, Corr{\^e}a, and
  Oliveira~Jr}{Fabbri et~al\mbox{.}}{2014}]%
        {mass}
\bibfield{author}{\bibinfo{person}{Renato Fabbri}, \bibinfo{person}{Vilson
  Vieira da~Silva Junior}, \bibinfo{person}{Ant{\^o}nio Carlos~Silvano
  Pessotti}, \bibinfo{person}{D{\'e}bora~Cristina Corr{\^e}a}, {and}
  \bibinfo{person}{Osvaldo~N Oliveira~Jr}.} \bibinfo{year}{2014}\natexlab{}.
\newblock \showarticletitle{Musical elements in the discrete-time
  representation of sound}.
\newblock \bibinfo{journal}{\emph{arXiv preprint arXiv:1412.6853}}
  (\bibinfo{year}{2014}).
\newblock


\bibitem[\protect\citeauthoryear{III}{III}{2018}]%
        {corpus}
\bibfield{author}{\bibinfo{person}{Jim~Henry III}.}
  \bibinfo{year}{2018}\natexlab{}.
\newblock \bibinfo{title}{lipu pi toki pona pi jan Jakopo}.
\newblock
\newblock
\urldef\tempurl%
\url{http://jimhenry.conlang.org/conlang/tokipona/tokipona.htm}
\showURL{%
Retrieved July 01, 2018 from \tempurl}


\bibitem[\protect\citeauthoryear{Knight}{Knight}{2017}]%
        {kama}
\bibfield{author}{\bibinfo{person}{Bryant Knight}.}
  \bibinfo{year}{2017}\natexlab{}.
\newblock \bibinfo{title}{o kama sona e toki pona!}
\newblock
\newblock
\urldef\tempurl%
\url{http://tokipona.net/tp/janpije/okamasona.php}
\showURL{%
Retrieved July 01, 2018 from \tempurl}


\bibitem[\protect\citeauthoryear{Knight}{Knight}{2018}]%
        {tpNetSH}
\bibfield{author}{\bibinfo{person}{Bryant Knight}.}
  \bibinfo{year}{2018}\natexlab{}.
\newblock \bibinfo{title}{Online gadget for syntax highlighting Toki Pona
  text}.
\newblock
\newblock
\urldef\tempurl%
\url{http://tokipona.net/tp/DisplayText.aspx}
\showURL{%
Retrieved July 01, 2018 from \tempurl}


\bibitem[\protect\citeauthoryear{Lang}{Lang}{2014}]%
        {tpLang}
\bibfield{author}{\bibinfo{person}{Sonja Lang}.}
  \bibinfo{year}{2014}\natexlab{}.
\newblock \bibinfo{booktitle}{\emph{Toki Pona: The language of good}}.
\newblock


\bibitem[\protect\citeauthoryear{Lang and Broholm}{Lang and Broholm}{2014}]%
        {interview}
\bibfield{author}{\bibinfo{person}{Sonja Lang} {and} \bibinfo{person}{Kris
  Broholm}.} \bibinfo{year}{2014}\natexlab{}.
\newblock \bibinfo{title}{AFP 20 – Sonja Lang: Toki Pona, Conlanging and the
  meaning of life}.
\newblock
\newblock
\urldef\tempurl%
\url{https://actualfluency.com/afp-20-sonja-lang-toki-pona-conlanging-meaning-life/}
\showURL{%
Retrieved July 01, 2018 from \tempurl}


\bibitem[\protect\citeauthoryear{Mehta and Zhu}{Mehta and Zhu}{2009}]%
        {blue}
\bibfield{author}{\bibinfo{person}{Ravi Mehta} {and}
  \bibinfo{person}{Rui~Juliet Zhu}.} \bibinfo{year}{2009}\natexlab{}.
\newblock \showarticletitle{Blue or red? Exploring the effect of color on
  cognitive task performances}.
\newblock \bibinfo{journal}{\emph{Science}} \bibinfo{volume}{323},
  \bibinfo{number}{5918} (\bibinfo{year}{2009}), \bibinfo{pages}{1226--1229}.
\newblock


\bibitem[\protect\citeauthoryear{Miller}{Miller}{1998}]%
        {wordnet}
\bibfield{author}{\bibinfo{person}{George Miller}.}
  \bibinfo{year}{1998}\natexlab{}.
\newblock \bibinfo{booktitle}{\emph{WordNet: An electronic lexical database}}.
\newblock \bibinfo{publisher}{MIT press}.
\newblock


\bibitem[\protect\citeauthoryear{Munzner}{Munzner}{2014}]%
        {dv1}
\bibfield{author}{\bibinfo{person}{Tamara Munzner}.}
  \bibinfo{year}{2014}\natexlab{}.
\newblock \bibinfo{booktitle}{\emph{Visualization analysis and design}}.
\newblock \bibinfo{publisher}{AK Peters/CRC Press}.
\newblock


\bibitem[\protect\citeauthoryear{O'Neill}{O'Neill}{2015}]%
        {sapWho}
\bibfield{author}{\bibinfo{person}{Sean~P O'Neill}.}
  \bibinfo{year}{2015}\natexlab{}.
\newblock \showarticletitle{Sapir--Whorf Hypothesis}.
\newblock \bibinfo{journal}{\emph{The International Encyclopedia of Language
  and Social Interaction}} (\bibinfo{year}{2015}), \bibinfo{pages}{1--10}.
\newblock


\bibitem[\protect\citeauthoryear{Schneider}{Schneider}{2018}]%
        {tp4}
\bibfield{author}{\bibinfo{person}{Stephan Schneider}.}
  \bibinfo{year}{2018}\natexlab{}.
\newblock \bibinfo{title}{nasin pi toki pona}.
\newblock
\newblock
\urldef\tempurl%
\url{https://github.com/stefichjo/toki-pona}
\showURL{%
Retrieved July 01, 2018 from \tempurl}


\bibitem[\protect\citeauthoryear{Schoonover}{Schoonover}{2011}]%
        {solarized}
\bibfield{author}{\bibinfo{person}{Ethan Schoonover}.}
  \bibinfo{year}{2011}\natexlab{}.
\newblock \bibinfo{title}{Solarized: precision colors for machines and people}.
\newblock
\newblock
\urldef\tempurl%
\url{http://ethanschoonover.com/solarized}
\showURL{%
Retrieved July 01, 2018 from \tempurl}


\bibitem[\protect\citeauthoryear{Spolsky}{Spolsky}{2008}]%
        {fonts}
\bibfield{author}{\bibinfo{person}{Avram~Joel Spolsky}.}
  \bibinfo{year}{2008}\natexlab{}.
\newblock \bibinfo{booktitle}{\emph{User interface design for programmers}}.
\newblock \bibinfo{publisher}{Apress}.
\newblock


\bibitem[\protect\citeauthoryear{Telea}{Telea}{2014}]%
        {dv2}
\bibfield{author}{\bibinfo{person}{Alexandru~C Telea}.}
  \bibinfo{year}{2014}\natexlab{}.
\newblock \bibinfo{booktitle}{\emph{Data visualization: principles and
  practice} (\bibinfo{edition}{2nd.} ed.)}.
\newblock \bibinfo{publisher}{AK Peters/CRC Press}.
\newblock


\bibitem[\protect\citeauthoryear{Viola, James, Schlangen, and Dijk}{Viola
  et~al\mbox{.}}{2008}]%
        {blue2}
\bibfield{author}{\bibinfo{person}{Antoine~U Viola}, \bibinfo{person}{Lynette~M
  James}, \bibinfo{person}{Luc~JM Schlangen}, {and} \bibinfo{person}{Derk-Jan
  Dijk}.} \bibinfo{year}{2008}\natexlab{}.
\newblock \showarticletitle{Blue-enriched white light in the workplace improves
  self-reported alertness, performance and sleep quality}.
\newblock \bibinfo{journal}{\emph{Scandinavian journal of work, environment \&
  health}} (\bibinfo{year}{2008}), \bibinfo{pages}{297--306}.
\newblock


\bibitem[\protect\citeauthoryear{Ward, Grinstein, and Keim}{Ward
  et~al\mbox{.}}{2015}]%
        {dv3}
\bibfield{author}{\bibinfo{person}{Matthew~O Ward}, \bibinfo{person}{Georges
  Grinstein}, {and} \bibinfo{person}{Daniel Keim}.}
  \bibinfo{year}{2015}\natexlab{}.
\newblock \bibinfo{booktitle}{\emph{Interactive data visualization:
  foundations, techniques, and applications}}.
\newblock \bibinfo{publisher}{AK Peters/CRC Press}.
\newblock


\bibitem[\protect\citeauthoryear{Ware}{Ware}{2012}]%
        {dv4}
\bibfield{author}{\bibinfo{person}{Colin Ware}.}
  \bibinfo{year}{2012}\natexlab{}.
\newblock \bibinfo{booktitle}{\emph{Information visualization: perception for
  design}}.
\newblock \bibinfo{publisher}{Elsevier}.
\newblock


\bibitem[\protect\citeauthoryear{{Wikipedia contributors}}{{Wikipedia
  contributors}}{2017}]%
        {wikiArtLang}
\bibfield{author}{\bibinfo{person}{{Wikipedia contributors}}.}
  \bibinfo{year}{2017}\natexlab{}.
\newblock \bibinfo{title}{Artistic language --- {Wikipedia}{,} The Free
  Encyclopedia}.
\newblock
\newblock
\urldef\tempurl%
\url{https://en.wikipedia.org/w/index.php?title=Artistic_language&oldid=808260717}
\showURL{%
\tempurl}
\newblock
\shownote{[Online; accessed 3-July-2018].}


\bibitem[\protect\citeauthoryear{{Wikipedia contributors}}{{Wikipedia
  contributors}}{2018a}]%
        {conlanWikip}
\bibfield{author}{\bibinfo{person}{{Wikipedia contributors}}.}
  \bibinfo{year}{2018}\natexlab{a}.
\newblock \bibinfo{title}{Constructed language --- {Wikipedia}{,} The Free
  Encyclopedia}.
\newblock
\newblock
\urldef\tempurl%
\url{https://en.wikipedia.org/w/index.php?title=Constructed_language&oldid=846512368}
\showURL{%
\tempurl}
\newblock
\shownote{[Online; accessed 1-July-2018].}


\bibitem[\protect\citeauthoryear{{Wikipedia contributors}}{{Wikipedia
  contributors}}{2018b}]%
        {wikiPoetry}
\bibfield{author}{\bibinfo{person}{{Wikipedia contributors}}.}
  \bibinfo{year}{2018}\natexlab{b}.
\newblock \bibinfo{title}{Poetry --- {Wikipedia}{,} The Free Encyclopedia}.
\newblock
\newblock
\urldef\tempurl%
\url{https://en.wikipedia.org/w/index.php?title=Poetry&oldid=847801129}
\showURL{%
\tempurl}
\newblock
\shownote{[Online; accessed 1-July-2018].}


\bibitem[\protect\citeauthoryear{{Wikipedia contributors}}{{Wikipedia
  contributors}}{2018c}]%
        {wikiToki}
\bibfield{author}{\bibinfo{person}{{Wikipedia contributors}}.}
  \bibinfo{year}{2018}\natexlab{c}.
\newblock \bibinfo{title}{Toki Pona --- {Wikipedia}{,} The Free Encyclopedia}.
\newblock
\newblock
\urldef\tempurl%
\url{https://en.wikipedia.org/w/index.php?title=Toki_Pona&oldid=846623837}
\showURL{%
\tempurl}
\newblock
\shownote{[Online; accessed 1-July-2018].}


\end{thebibliography}

\end{document}